# Classification of the factors of functioning of the individual human thermoregulation system and prediction of its state (local thermophysical aspects)

Dmitry Simankov

https://orcid.org/0000-0001-7970-5422

https://www.elibrary.ru/author_profile.asp?authorid=678324

Moscow Aviation Institute (National Research University), Moscow (https://mai.ru)

Postal address: Russian Federation (RU), 125993, Moscow, Volokolamskoe highway, 4,

(Institute No. 8)

pegasds1@mail.ru

**Abstract.**

A classification of the tasks of human thermoregulation is proposed, which allows combining various scientific studies into one system, which takes into account environmental factors, biological feedback in the body and different properties of the constituent parts of the human thermoregulation system. The proposed multifactorial model of the human thermoregulation system is based on combining the classification of tasks represented by a group of main factors (temperature, time and the characteristic size of the system under consideration), with the classification of tasks in the studied sciences.

Applying an interdisciplinary approach to the study of human thermoregulation (thermogenesis), it is possible to model complex systems that allow predicting the state of the human thermoregulation system. A multifactor model, based on the systematization of knowledge, can be used as the basis for computer modeling of a human thermoregulation system.

Classification of tasks for the study of the human thermoregulation system allows you to systematize scientific knowledge and identify unexplored aspects for future research. Research and modeling of the human thermoregulation system are the scientific basis for the creation of new measuring, diagnostic and therapeutic equipment, new treatment methods and, possibly, a standardized personalized calculation method in the form of a software and hardware complex with an online calculator via the Internet of the influence of external factors (weather) for predicting homeostasis (and the state of the thermoregulation system) of a person.

The article discusses a number of the most significant scientific works for each area of research. Identified and systematized the most important problems for study in each of them from the point of view of heat and mass transfer and energy conversion at different levels of the description of biosystems in the human body.

**Key words**: skin temperature, thermal comfort, human thermal receptors, thermoregulation system, mathematical modeling, skin thermomechanics, bioheat transfer.

**1. INTRODUCTION**.

From the general course of mammalian physiology, human thermoregulation is understood as the body's ability to maintain a constant core temperature under different external environmental conditions. The function of the thermoregulatory system is to maintain and stabilize a constant body temperature, thanks to which the body can exist and develop. In the process of life, a person is exposed to a wide range of temperatures in which his thermoregulation system functions. This is due to work (firefighters, steelworkers), medical treatment (cryosurgery, radiotherapy), climatic zone of residence (people of the far north, aborigines of tropical rain forests, nomads of sandy deserts) and so on.

The human thermoregulation system is one of the systems of the human body that ensures its existence in conjunction with other body systems. All processes (physical, chemical, physiological, biological) in a person are interconnected and interact with each other at different levels of the biological system. To describe any system in the human body, it is necessary to establish all its interconnections with other systems. The biological system (respiratory, circulatory, thermoregulation, sexual, and so on) can be described in terms of functioning by different sciences (physicochemical, biological, biophysical, and so on) on different (quantum, subcellular, cellular, tissue, organ, the whole organism, population ) levels. Now there is no general model of thermoregulation of a living organism, described by one system of equations taking into account several interrelated levels, due to its complexity for computer modeling, since the human body is a complex and interconnected system that has yet to be fully understood. The system of thermoregulation of the human body, as a complex problem, is solved by a deductive approach, that is, from the general to the particular (by its constituent parts), since the level of measuring technology, mathematical methods and means of calculations are developing gradually.

In case of malfunction of (part) of the human thermoregulation system in medicine, it is diagnosed as a disease: hypothermia (ICD-10 - T68) - a state of the body in which the body temperature drops below that required to maintain normal metabolism and functioning; hyperthermia (ICD-10 - T67) - overheating, accumulation of excess heat in the human body with an increase in body temperature, caused by external factors that impede heat transfer to the external environment or increase the flow of heat from the outside. In addition to the ICD classification, there is another approach in terminology for thermal injuries, described in works [177, 192].

The human thermoregulation system can function normally and abnormally (before / after illness and during illness). The normal mode of functioning of the human thermoregulation system is more clearly described in terms of labor protection and is characterized as a comfortable state of the body in the environment. The comfortable state of the body is understood as a combination of external environmental factors and the psycho-physiological state of a person in which the processes of vital activity in the body proceed normally. This is a subjective feeling for every person. ISO 7730: 2005 ("Ergonomics of the thermal environment - Analytical determination and interpretation of thermal comfort using calculation of the PMV and PPD indices and local thermal comfort criteria ", IDT). For local short-term exposures at different temperatures, there are methods for assessing human response - ISO 13732-1: 2008, ISO / TS 13732-2: 2001, ISO 13732-3: 2008. In addition, there is a number of scientific and regulatory works to provide workers with a safe working environment.

This work is aimed at popularizing, updating and supplementing with fresh information about the system of individual thermoregulation for work "Protecting workers from a heating microclimate" [1] from the US Department of Health and the "Encyclopedia of Occupational Safety" from the UN ILO [13]. Protecting Workers from Warming Climate Assessments was carried out by the US Department of Health in 2016 in conjunction with the Centers for Disease Control and Prevention and the National Institute for Occupational Safety and Health. It is a document for updating over time, available in the public domain in English and translated into national languages in several countries, where it is also in the public domain and can be used. A similar work was carried out by the International Labor Organization at the United Nations (ILO is a specialized agency of the United Nations) and published in the public domain in English and French. The "Encyclopedia of Occupational Safety" from the UN ILO is also being updated and the 4th edition is now available [13]. For a review of the thermoregulation system, see «Recent advances in thermoregulation» [230] and the 2018 book «The thermoregulation system and how it works» [232]. The book covers the physiology and neuroanatomy of the thermoregulation system and provides descriptions of how the regulation of body temperature interacts with other physiological functions (such as sleep, osmoregulation, and immunity), stress, exercise and aging.

Many works have been devoted to human thermal interaction with the environment. In these works, the prediction of the thermal reaction of the human body or its parts to various environmental factors is considered. Thermophysiological models of recent years determine not only the heat transfer between the human body and the environment, but also the biological feedback inside the body to environmental factors. Thermophysiological models consist of passive and active elements. In the passive thermal model, the heat balance equation is written in different approaches [58]. The active thermal model takes into account the cold / heat receptors that regulate the thermoregulatory process by narrowing and widening the blood vessels of the skin. A part of the human's active thermoregulatory system is shivering (heat generation from muscle contraction), the process of sweating (heat loss due to evaporation), heat generation from the fatty layer of the hypodermis skin (non-shivering thermogenesis). In a state of comfort, the active thermoregulation system "controls" the passive system [4]. However, many physiological and biophysical aspects are not taken into account in the model of an active thermoregulatory system.

To develop a predictive model of a human thermoregulation system, a lot of experimental and theoretical work will be required to obtain an adequate model system of equations and a method for calculating the state of this body system. To do this, it is necessary to break down the human thermoregulation system into its component parts. To study external and internal factors influencing the functioning of its parts at different levels of the description of biological systems, to establish biological feedbacks and interactions in the body, which are either not clearly observed or have not yet been described mathematically.

The identified needs in the study of biological systems will push to the creation of techniques and technologies for future research and to unite researchers from different fields of science. To systematize the directions of studying biological material at different levels of its organization (tissue, cellular), you will need logic in construction and a tool for visualization and systematization. A multidimensional tabular form appears to be the most suitable tool for systematizing and visualizing factors and tasks in the field of human thermoregulation when describing it at different levels of biomaterial (tissues, cells) organization. An alternative to multidimensional tables is the tree method (area of knowledge - statistics), built according to a specific algorithm. But this will complicate the logic of understanding the interactions in the body and will not reveal the relationship in the complex system of thermoregulation.

## 1.1 Practical application: now and in the future.

The benefit of studying the factors of the human thermoregulation system is the creation of a mathematical model of the body system, which makes it possible to diagnose and predict its state. An indirect benefit from considering the factors of the human thermoregulation system is the pooling of knowledge in the chosen field to search for new research and the unification of various people of researchers.

Two main directions for the practical application of the classification of factors of the human thermoregulation system are highlighted: the field of labor protection (regulatory documents) and medicine. The following reasoning is indirectly related to the considered classification of thermoregulation factors and is more related to future directions of development in this topic: analytical system, equipment, standardized and personalized assessment and diagnostic methods, treatment methods, recommendations on labor protection, rules for recruiting for work and service, and more.

When systematizing the accumulated knowledge in the field of human thermoregulation and setting the goal of computer modeling of this body system for its personalized application and consolidation in regulatory documents in the field of occupational safety and health, the approach with multidimensional tables will identify problems that have not yet been solved, and justify the need for their solution for achieving a common goal.

Labor protection and research into the capabilities of the body are now becoming even more relevant in connection with the development of space tourism, long distance flights in space, as well as in the development of the Arctic and Antarctica and other areas of human activity. In extreme environmental conditions, the tasks of protecting a person and the correct selection of people for certain types of work become relevant, which can be more correctly solved using information on the thermoregulation system and its individual diagnostics when using predictive models when designing clothing for people in space, firefighters, steelworkers, military personnel and others.

The traditional demand for new information on the functioning of the body remains in sports medicine, disaster medicine, and military medicine. More on this below. With the help of modern Big Data and AI technologies, in relation to solving a large complex problem (human thermoregulation system), it will be possible to combine particular subtasks and find unidentified relationships and their dependencies. This will help diagnose some diseases in people and adjust the treatment plan, taking into account the individual characteristics of the patient's body.

In addition, for the medicine of the future, other areas can be identified that can be developed and improved as applied problems. The classification of factors of the human thermoregulation system will partially help solve such problems in the form of a tool to identify new areas for research and combine knowledge from different fields of science, helping different researchers to better understand the relationships in the human body.

A number of applied problems in the field of medicine can be distinguished, where knowledge about thermophysical aspects and the system of human thermoregulation is used.

1) In cryo-operations, doctors calculate the necrosis zone (the dynamics of the growth of the demarcation zone) based on their experience. The existing empirical maps of the freezing zone for apparatuses are approximate. There is no exact calculation of the zone of freezing and necrosis, and during freezing healthy tissue may be injured, and the malignant pathological tissue may not freeze enough to die. There is no cryo surgery planning system (there have been various attempts to create it).
2) For the treatment of hypothermia and hyperthermia, a pharmacological approach is used, and the diagnosis is based on the external signs of the patient. It is now impossible to determine the individual threshold and conditions under which this disease will be in humans.
3) For sports medicine, it is necessary to be able to assess and diagnose the thermoregulation system of athletes in order to more fully measure the state of the athletes' body and its changes during training. This is necessary so that exercise does not harm your health. Most professional athletes increase their body indicators during training at the expense of others, which negatively affects the future, after a sports career, and in old age.
4) In disaster medicine, when a person is diagnosed with extensive thermal injuries from high or low temperatures, doctors make diagnostics visually and rarely measure pressure and heart rate without contact. In such cases, a blood test can not always be quickly performed in a deployed manner to assess the internal state of the body. Physiotherapy and pharmaceutical treatments are used. The effectiveness of these methods depends on the drugs used. So, with the optimal combined intake of drugs, for a hypothermic body, you can quickly and safely activate internal sources of heat in the form of the work of organs, and force the heart to more actively pump blood with components for various internal organs that will help save a person.

5) For military medicine, tasks in the field of human thermoregulation are interesting - providing a serviceman with means that allow him to cope with low ambient temperatures. There are hand and face creams, a different diet. Sometimes alcohol-containing foods are used in conjunction with high-energy foods. From a pharmaceutical point of view, the composition of human blood and its circulation rate can be regulated, which makes it possible to influence the thermoregulation system. In military affairs, you can also use a diagnostic system to assess the individual thermoregulation of a person and his sensitivity and tolerance to cold or heat in order to compose individual and collective physical exercises. Now this is not taken into account when training soldiers.

6) In radiology, when planning an operation, the thermophysical properties of biological tissues are used, usually in the form of constants or with a dependence on temperature. The thermophysical properties of biological tissues have the properties of anisotropy, which is not taken into account in calculations. It is important that each person has the physical properties of biological tissues are individual and changeable due to external and internal reasons. This is not taken into account anywhere in medical practice, since there is no measuring equipment and is not used. In the presence of such diagnostic equipment (and the regularities of the influence of factors on the properties of biological tissues), it is possible to plan an operation in radiology and cryosurgery more accurately.

7) In space medicine (during long flights, in the absence of gravity or gravity of another planet or satellite), the human thermoregulation system, like his other body system, becomes a subject for study so that people can survive and adapt in such conditions.

8) The climate on the planet is changing - it is getting warmer, and humanity is adapting to climate change. The human body begins to function differently, trying to adapt, including its thermoregulation system. Each person during his life the rate of adaptation is different due to a number of factors, but now this is not noticeable due to the relatively small change in temperature on the planet upward. However, the dynamics of temperature growth on the planet is only increasing, which means that there will be a time when the rate of adaptation of the organism will be insufficient to adapt to the new environment. The groups of people who were able to quickly adapt and who did not will become more contrasting and clearly distinguished. For the latter group of people, medical pharmaceuticals may be required to change the genotype and strengthen some human organs. Research in this area may require a diagnostic system and additional equipment for personalized research and the manufacture of personalized pharmaceuticals. Including knowledge and equipment for diagnostics of the human thermoregulation system.

9) Now they make various biological and mechanical artificial organs, which are sources of heat in the body. Therefore, it is important to have complete information about different systems of the body so that artificial and mechanical organs and parts of the human body work correctly. The same applies to wearable electronics (chips).

In routine clinical practice, examination does not assess the human thermoregulatory system. Usually, the respiratory, circulatory, lymphatic, reproductive and other systems of the human body are assessed, but not the thermoregulation system. However, even if all organs work in the parameters acceptable for them, this does not mean that any of the body systems is working in optimal parameters. We need a detailed diagnosis in order to timely prevent the onset and development of the disease, for example, hypothermia and hyperthermia. Usually, a neurologist can diagnose a violation in the functioning of the human thermoregulation system.

To date, a lot of monitoring and diagnostic equipment and research methods have been created. They continue to develop and improve, new ones appear. If there is a hardware and software complex with additional measuring equipment for diagnosing the human thermoregulation system, it will be possible to diagnose other diseases, in addition to hypothermia and hyperthermia.

In the future, it is possible to create a new generation of hardware and software complexes and the principle of action for studying the systems of the body. As a result, this will lead to the diagnosis of the organism at the cellular level, which is currently unattainable due to limitations in the computing power of technology, the equipment used and the lack of some knowledge. <u>The proposed classification of factors of the human thermoregulation system is intended to systematize and highlight these gaps in knowledge.</u> The accumulated knowledge in the future will make it possible to create a program (computer model) of the state of the systems of the human body, at first as separate, and in the future of the whole organism at the cellular level, which now looks like fantasy.

**1.2 Review of general mathematical models and observations, which now partially reflect the human thermoregulation system and heat and mass transfer processes at different levels of description of biological systems.**

Let us consider the thermophysical aspects in their general form, and will be described in more detail below.

1) <u>A group of people</u>. They mainly use statistical methods for research.
    1.1) For a group of people and even the population of a city or region in many countries, studies have been and are being carried out on the average body temperature in different anatomical sites depending on gender and age. And also for such groups of people, they explore the dynamics of the zone of comfortable sensations under the influence of climate change and living conditions.
    1.2) In sports medicine, comparative analyzes of the indicators of the functioning of the athlete's body are made with the average figures of a healthy person, taken from the literature. For example, the content of subcutaneous fat, the volume of exhaled air, the energy of muscle contraction, body surface temperature, etc.
    1.3) The study of human metabolism was carried out in different age groups, during sleep and wakefulness, fluctuations were established throughout the day.
2) <u>The human body</u>. Part of the work, especially in physiology, was carried out by statistical methods. For the last quarter of a century, various empirical and semi-analytical models have been actively used, including the use of computer modeling, describing the heat-physiological and energy processes of the body.
    2.1) Many heat-related physiological models have been created for the human body, of which about 8 are considered relevant. Polysegmental and neurophysiological models with physiological feedback with modeling of the body's responses to receptors in the skin for thermoregulation of the human body appear.
    2.2) To simulate the thermal effect of the environment on a person, various approaches have been developed for calculating the effective exposure temperature, which take into account air movement, humidity, atmospheric pressure, and thermal radiation. The most universal is the indicator of the complex environmental impact on the whole human body - UTCI, which was created by a group of organizations on a large sample of people, and its peculiarity is that it predicts the behavior of the state of the body (physiological reactions of the body) to changes in environmental parameters [19].
    2.3) The mechanisms of human adaptation to cold and hot constant environmental conditions are studied. The physiological reactions of the body are studied by comparative analysis.

2.4) There is an estimate of the body's energy expenditures for performing a certain work, as well as empirical dependences of different indices, for example, the ratio of body weight to skin area, have been established.
3) <u>An organ or part of the body</u>. In modeling, three-dimensional objects are used, including real geometry from medical images (DICOM files), and numerical modeling is often performed in COMSOL Multiphysics, ANSYS with connected special libraries for numerical calculation methods. Heat and mass transfer processes, energy, are modeled, phase transformations and radiative heat transfer are taken into account. Sometimes the phenomena of osmosis, pressure and velocity in blood vessels, air velocity in the lungs and on the surface of the skin are taken into account.
   3.1) Radiology calculates the radiation dose to the tumor and a safe way to supply such energy so as not to harm healthy tissues.
   3.2) In PET (SPECT) studies, areas of location and volume of tissue that have ceased to function are studied.
   3.3) Older human skin models that did not take into account its features were used to assess heat loss or thermal / cold injury.
   3.4) Breast cancer can be diagnosed both by the thermophysical properties of the skin (increased coefficient of thermal conductivity), and by thermometric methods with the localization of the lesion and an assessment of its volume. The tumor has an increased cell metabolism, which contributes to the total heat flow in a local area of the human body.
   3.5) Muscle energy is a factor in the heat production system during movement. This is an additional heat flow during physical exertion, which comes from the muscles during their contractions..
   3.6) Thermal imagers and thermal imaging systems by the method of microwave radiothermometry record infrared radiation of the temperature background of the human body, which is invisible to the human eye, which reflects the pathological processes occurring in the body. Microwave radiothermometry is used for early diagnosis of a tumor [188]. During the COVID-19 pandemic, individual and streaming measurements of human temperature have increased to identify people with possible illness. Previously, they were popular in the diagnosis of cancer, but due to the characteristics of the body (nervous system) of the patient, they could give an incorrect examination result and are now used as auxiliary means of diagnosing various diseases, giving indirect information about disorders of the functioning of internal organs by their thermal field.
   3.7) Thermal stimulation of the congestion of nerve endings. You can act on the throat from the inside in a certain place with a cold object in order to cause cold stimulation of the body's immune system to fight viral diseases (Kochenov V.I.) [169]. The method of pulsed heat exposure (thermal pulsation) on the reflexogenic zone of the face has proven itself well in the treatment of neuroses and neurosis-like states of various nature. (Timoshenko D.A. and D.D.) [103].
   3.8) Cold stimulation of the human body surface in order to activate the immune system is used in cryosauna procedures. Thermal stimulation of the surface of the human body in the bath is aimed at warming up the body and accelerating the course of internal processes that stimulate various physiological systems. Conditions such as a bath (prolonged heat exposure) or cryosauna (short cold exposure) stress the body and make it activate the internal mechanisms of survival in extreme conditions, which can favorably affect the general state of the human body after the procedure.
4) <u>Organ tissues</u>.
   4.1) Modeling the liver and lungs as a porous material is necessary when studying the absorption coefficient of gases and liquids to assess their condition and diagnose the disease.

- 4.2) Cryodestruction of tumors of different localization is performed with cryodestructors of different geometric shapes and cooling rates. A microwave field is often used for faster and deeper freezing of large volumes of pathological malignant tumors. Less commonly, special ointments (cryoprotectants) are used. Most cryo-operations are done precisely due to the experience of the doctor, who in practice knows the approximate zone of necrosis.
- 4.3) Thermal ablation is a widely used method for treating tumors in organs such as liver, kidneys, etc. It is an alternative method of cryodestruction, where you also need to know the modes of heating thermodes and heating biological tissues. [184, 202]
- 4.4) Thermal stimulation of targeted nerve circuits using remotely controlled nanotransducers: therapy for neurodegenerative disorders. (Neurodegenerative disorders, such as Alzheimer's and Parkinson's, are signal abnormalities in neuronal cells. Deep brain stimulation (DBS) is the traditional treatment for Parkinson's disease.) [170]
- 4.5) Modern models of human skin describe well the models of bioheat and the body's response to heat / cold local effects on receptors.
- 4.6) The behavior of anastomosis (arteriovenous shunts) depends on the environmental conditions to which the central nervous system responds.

5) <u>Cells of biological tissue</u>. At the cellular level, it is advisable to describe the conducting nervous system of the body, along which the signals for controlling the thermoregulation system go. As well as the mechanisms of humoral and hormonal thermoregulation. When describing systems, biophysical methods are used.
   - 5.1) Heat and cold receptors are located in the skin, internal organs and blood vessels that respond to temperature changes and send signals to the central nervous system (and the spinal cord). It is believed that a 0.1 K change in the temperature at the receptor causes a change in the signal, but this is convenient for modeling.
   - 5.2) From receptors, signals travel to the brain. The main part of the brain responsible for thermoregulation is the hypothalamus. Signals from receptors are processed not only by the hypothalamus, but also by other parts of the brain, the cerebral cortex and the spinal cord. This means that the region of the brain that receives the signal does not necessarily process it and produces a return signal.
   - 5.3) Outgoing signals from the brain in the thermoregulation system can be divided into behavioral (conscious mechanisms in the somatic nervous system) and internal nervous ones in the form of the sympathetic and parasympathetic nervous systems (autonomic nervous system) to regulate muscle tone and heart function. These signals trigger the internal and external mechanisms of heat production and heat transfer. The main tasks are the study of biochemical reactions, their interrelationships, identification of signals, their characteristics, methods of signal processing in the brain, and, as a consequence, different energy processes in different parts of the body. [171]
   - 5.4) The hormonal mechanism of human thermoregulation uses hormones from the pituitary, thyroid and adrenal glands. These organs produce various hormones that, at low ambient temperatures, enhance oxidative processes in tissues, in particular in muscles, increase heat production, constrict skin vessels, thereby reducing heat transfer; at high ambient temperatures affect the arterioles of the skin, promote their vasodilation, thus participating in the processes of heat transfer. The main tasks are the study of biochemical reactions and their relationship. [171]
   - 5.5) The humoral (neurohumoral) system of thermoregulation occurs due to mechanisms regulated by the sympathetic nervous system (oxidative phosphorylation, glycogenolysis, glycolysis in the liver and lipolysis in brown fat) and by the somatic nervous system (regulates contractile muscle thermogenesis). The main tasks are the study of biochemical reactions, their relationship and the energy of muscle contraction. Not all of the generated energy is converted into macroenergetic bonds, and in these reactions more is converted into heat. [171]

6) <u>Molecules of biological tissue cells</u>. [37]

6.1) Hydronium ions $H_3O+$ in glucose injections into the tumor during irradiation promote a more rapid accumulation of energy and, as a result, the destruction of tumor cells occurs.

6.2) The electromagnetic field in a computational experiment using the molecular dynamics method has shown that it can change (reorient) the dipole moment of protein molecules and thereby change its secondary structure, which contributes to denaturation.

6.3) When a living cell is frozen, water molecules can be incorporated into amino acid residues (complex proteins). And for deep freezing of proteins, cases are possible when free water does not freeze in the intercellular fluid, probably due to the balancing of van der Waals forces and hydrogen bonds between molecules in the crystal.

6.4) Depending on the rate of cooling, the mechanisms of rupture of the cell membrane (ice crystals from the cell or intercellular fluid) are distinguished.

6.5) The mechanisms of cell destruction are apoptosis (normal, programmable) and necrosis from temperature (physical or chemical effects).

6.6) The breaking of hydrogen bonds prevents the nucleation of ice crystals, which lowers the crystallization temperature. This reduces the number of water bonds in cells and increases the thermal conductivity of the medium. This is used in cryosurgery as a mechanism to increase the volume of cryodestruction of pathological biological tissues.

6.7) The role of free water molecules in globular proteins is very important. Water on biological surfaces rearranges hydrogen bonds and exhibits a number of specific properties. Biological water can be divided into 3 types: free, bound and having one hydrogen bond with polar groups of macromolecules.

6.8) A decrease in temperature triggers the synthesis of "heat shock proteins" (sometimes called heat stress proteins and cold shock proteins), which maintain cell homeostasis (compensate for a decrease in the rate of metabolic reactions).

6.9) The mechanisms of protein denaturation under the influence of high and low temperatures.

6.10) Properties of cell membranes: permeability, osmosis, potentials, extended liquid mosaic model in modeling, phase transitions, the effect of specific proteins on the state of the membrane, nucleation, dehydration of membranes and cells, membrane lysis (destruction), activation energy, simulated cell shape, mechanisms of ice formation and its melting inside and outside the cell, the influence of cryoprotectants (substances).

Many computer programs for visualization of human organs, blood vessels and nerve fibers and other structures have been created. The most famous are such programs for 3D visualization: 3D-Body Adventure (USA), Advantage Windows (USA), ADAM (UK), Corps Human (France), Body Voyage (USA), DUCT (UK). [172]

Modeling of individual human organs has recently become a trend, including the thermophysical state in health and disease. A huge number of bio-heat models have been created for the skin, among the latter are combined models that take into account radiation heat transfer, anisotropy of the medium, convective flows of mass and energy inside the medium, as well as bio-heat models take into account the reverse biological reaction by receptors.

Mathematical models of the lungs and liver, as porous media, and other human organs have been created in the computer programs COMSOL Multiphysics, ANSYS and others to plan their treatment. Thanks to the diagnostic technique, it is possible to visualize the bloodstream of the whole organism, including the capillary bed in the internal organs, as well as visualize hard and soft biological tissues, which is especially necessary for radiological planning of surgery.

## 1.3 Criticism of existing mathematical models of thermoregulation systems and thermophysical aspects of a person.

The main disadvantages and weaknesses in modeling heat and mass transfer processes for the human body can be grouped according to the level of description of biological systems. Any mathematical model has strengths and weaknesses. The more complex the mathematical model, the more unknown variables and coefficients in it, as well as the presence of relationships between variables. General disadvantages of biological mathematical models describing real processes at different levels are: the accuracy (uncertainty) of measurements is higher than 3%, the reverse biological factors and the presence of diseases are little taken into account, data verification is practically impossible, there is no personalized approach (a set of specific factors for one person).

In accordance with the given thermophysical aspects for humans, the disadvantages of the models can be listed.

1. A group of people. The individual characteristics of a particular person and his disease are not taken into account. There is no large sample for statistical analysis. The peculiarities of groups of people living in different climatic zones or employed in the same profession are not taken into account. It is not correct to make statistical models on Europeans and then give these data for practical use for residents of the Asian region, since people live in different climatic zones. It is not correct to make generalizations from an incorrectly formed sample of people, since such people have many biological characteristics that are generally not taken into account in the standards for labor protection and ensuring comfortable conditions for a person in the workplace. A number of scientific articles confirm this obvious fact. For example, in [23] it is shown that the statistical model PMV (or PPD approach) for the Chinese, when replicating the experience, gives different generalized numerical values in the design engineering formulas than for the Europeans. And the division of people by race is not correct from the point of view of describing physiology or genotype, it is often politicized. Division by race is a social construction that has nothing to do with the natural sciences [137].
2. The human body. It is often modeled as a multi-layer cylinder or a group of cylinders that does not reflect the real geometry of the human body. Biotissue coefficients are universal and are constant values, which does not correspond to reality. Physiological factors are practically absent everywhere. In empirical formulas, correction factors are used that take into account one of the physiological factors, but not their total complex (effect) and the dynamics of change over time. In empirical formulas, correction factors are used that take into account one of the environmental factors, but not their total complex (effect) and the dynamics of change over time.
3. An organ or part of the body. While many models have been created for the skin that take into account many physical and biological factors, including combined models, this has not been done for other organs. If in CAD systems approximate models of different morphologies of internal organs are created, then there are still no physical (and physiological) models that take into account the functioning of organs in the body with biological feedback (receptors, hormones, the nervous system). Operation planning systems (cryo-, surgery, radiology, thermal ablation, ultrasound) are just developing (except for radiology, where medical physicists have long been calculating the radiation dose in computer programs).
4. Organ tissues. The porosity and anisotropy of biological tissues, their nonlinear dependence of properties on temperature, mechanisms with vascular reactions are not always taken into account; change in heat production under the influence of human behavior (behavioral conscious mechanism of thermoregulation), under the influence of hormonal and humoral thermoregulation or other physiological mechanisms of thermoregulation in humans.

5. <u>Cells of biological tissue</u>. During freezing and thawing, it is difficult to determine the state of ice and its transitions to other states (now 20 types of ice have been discovered). Different cryoprotectants (by injection or ointment) interact in different ways with the cell membranes of different organs and blood vessels of the bloodstream. Redox reactions proceed differently in different organs due to the presence of a number of factors. It is not possible to measure specific parameters of the environment in a specific person at the moment (the coefficients are taken from experiments in simulated environments and from computer modeling using molecular dynamics methods). Each cell has a life cycle and it is not yet possible to assess the composition and its components by cell age to take this into account in cell models. Modeling the ice front and its composition for an ensemble of cells does not take into account macroeffects and the energy of phase transformation of one type of ice into another. Computer modeling uses different potentials on the cell surface, but there is no common understanding and transition of models from a small characteristic scale with a short process time to larger models with sizes and times, taking into account energy fluctuations in large systems (a group of cells).
6. <u>Molecules of biological tissue cells</u>. Protein denaturation under the influence of temperature occurs in several stages, and it is now problematic to estimate the exact values of the amount of proteins that have survived, partially converted or destroyed. The processes of ATP formation are of a stochastic nature, and during the decay of ATP, it is difficult to know exactly what energy has gone into heat and what energy has gone into macroenergetic bonds. When modeling proteins on a computer, it is not always possible to take into account all the long-range and short-range forces that keep the molecules in the same chain with the tails. With different physical processes (phase transitions of ice into different states during freezing and thawing; during thermal burns; when pumping energy from ionizing radiation or ultrasound), it is not always possible to determine: a part of the energy that untwists the main protein filament and breaks the tails in this chain (macroenergetic communication); part of the energy that breaks down protein chains; part of the energy that converts into heat (radiant / radiative heat transfer); part of the energy that is converted into macroenergetic bonds during redox reactions and during the synthesis or rearrangement of macromolecules; part of the energy that is converted into photons. The coefficients in such complex mathematical models are selected from experiments "in vitro", either indirectly, or by methods of comparative analysis, which, although difficult, can differ by significant values in a real experiment. The motion of electrons and hydroxyl groups in modeling is very important, but for each specific particle or portion of energy (when simulating systems using quantum mechanics methods) it is impossible to trace, and these processes are described by operators and statistical observable functions. But so far there are too few operators for the transition from microscopic models to more complex ones so that the observed phenomena in macroscopic models would be taken into account.

**2. Formulation of the problem.**

It is proposed to divide the human thermoregulation system described in physiology into parts and analyze it at different levels. Using the inductive approach when classifying the factors of thermophysical aspects, show the places where additional research is needed.

**OBJECTIVES OF THE WORK**:

1. Combine information on thermophysical aspects of a person from different branches of science, especially the system of human thermoregulation.
2. For the system of standardization of different countries to show the need to formulate, measure, calculate, select criteria and apply environmental (climate) data differently.
3. Show specialists from different fields of science different biological factors when modeling temperature (and other) processes.

4. To propose a classification system for thermophysical problems for humans for the future further multifactorial model for predicting the state of an organism or its part.

**Methodology.**

The search for scientific publications was carried out in the following scientific data bases on the Internet: Science Direct (Scopus), Web of Science (cel.webofknowledge.com - no), PubMed, Researchgate, and eLIBRARY.RU. Different combinations of words were used. The search was conducted by reviewing journals written in English on human research. The search included literature that the author had encountered earlier (the method of "personal knowledge" or academic and personal contacts). "Accidental discoveries" occurred while studying patents and doctoral dissertations. In addition, the list of references includes works available in the Lenin Library in Moscow Russia. A feature of the current state of scientific bases is that one and the same journal is indexed in several bases at once, and this facilitates the search for scientific papers.

The main criterion by which the works included in this article were selected was their completeness of the review on the selected topic (review articles) or specific works that are rarely mentioned, but are directly related to the topic under study, capable of significantly improving the understanding the heat balance of a person and its extreme values.

An additional criterion was the novelty of the articles, especially from 2016. Articles should cover the whole range of temperatures - from cryogenic to burns (ICD 10: T20-T25, T27, T29-T31, L55) and describe phenomena at different levels of complexity by different research methods, but mainly by engineering and physical techniques with mathematical models and dependencies. For tissues and cells, in vivo studies were chosen, and in vitro studies were not generally considered.

**2.1 Technical part. Factor classification system.**

Thermophysical aspects of thermoregulation and other thermal phenomena in the human body are studied and described in various sciences - general biology, physiology, biophysics and biochemistry. In the classical sense, any system has its own characteristic size (L) and process time (t), from the point of view of physical description. The article deals with thermophysical aspects in the human body, in particular, its thermoregulation system. Therefore, temperature (T) will be included in the main parameters in the mathematical description of the human body system. In order to systematize various factors of the external environment and internal processes, it is proposed to make a division according to the principle from a larger biosystem to a smaller one, namely, to single out such groups with the symbol "S": a group of people (S1), the body of one person (S2), part of the human body (S3), tissue (S4), cell (S5) and molecule (S6). Each of this group has its own scale (L), process time (t) and characteristic temperature (T). Schematically, the systematization logic can be presented in the form of Table 1.

Table 1. Characteristic parameters of biosystems (S, L, t, T)

|   | Biosystem size (S) | Branch of science | Typical system size (L), $L_1 < L_2 < L_3$ | The characteristic time of the process in the system (t), $t_1 < t_2 < t_3$ | Temperature divisions (T), $T_1 < T_2 < T_3$ |
|---|---|---|---|---|---|
| 1 | Group of people (S1) | General biology. Statistics. | L1<1m this is S3<br>L2=1-2,2m<br>L3>2,2m this is Ø | t1<1 hour this is S2<br>t2=1-24 hours<br>t3>24 hours | T1<32°C<br>T2=32-37°C<br>T3>37°C |
| 2 | Human body (S2) | General biology. Physiology. Statistics and Bioengineering. | L1<1m это S3<br>L2=1-2,2m<br>L3>2,2m this is Ø | t1<5 min<br>t2=5 min -12 hours<br>t3>12 hours this is S1 | T1<32°C<br>T2=32-37°C<br>T3>37°C |

| 3 | Body part (organ) (S3) | Physiology. Medical physics. | L1<3cm this is S4<br>L2=3-100cm<br>L3>100 cm this is S2 | t1<5 min this is S4<br>t2=5 min -12 hours<br>t3>12 hours this is S2 или S1 | T1<0°C this is S4<br>T2=0-45°C<br>T3>45°C this is S4 |
| --- | --- | --- | --- | --- | --- |
| 4 | Fabric (S4) | Physiology. Biophysics. Medical physics. | L1<0,005mm this is S5<br>L2=0,005-30mm<br>L3>100mm this is S3 | t1<0,5 min this is S5<br>t2=0,5-30 min<br>t3>30 min this is S3 | T1<0°C<br>T2=0-45°C<br>T3>45°C |
| 5 | Cage (S5) | Biophysics. Biochemistry. Medical physics. | L1<0,005mm this is S6<br>L2=0,005mm<br>L3>0,005mm this is S4 | t1<1s this is S6<br>t2=1-30s<br>t3>30s this is S4 | T1<34°C<br>T2=34-38°C<br>T3>38°C this is S6 |
| 6 | Molecule (S6) | Biophysics. Biochemistry. Quantum Physics and Chemistry (QPC). | L1<$10^{-10}$m this is QPC<br>L2=$10^{-10}$m<br>L3>$10^{-10}$m this is S5 | t1<$10^{-12}$s this is QPC<br>t2=$10^{-12}$-1s<br>t3>1s this is S5 | T1<0°C<br>T2=0-45°C<br>T3>45°C |

Each figure has its own rationale in Table 1. There are no strict rules and it is necessary to understand the general logic when compiling this table. An explanation of each parameter is provided here. For a group of people (S1), the typical size is the average height of a person, including children. The Ø sign is an empty set, or "dead end" (Russian word) of the study. If the object of research in S1 and S2 is less than 1 meter, then this is already a part of the body and it should be considered in research S3, that is, when describing the object, models and description of its functioning are already used in terms of the field of science that corresponds to its size. If a group of people is studied for less than 1 hour, then it is more correct to describe physiological changes, rather than adaptive mechanisms to climate change, which are developed and observed for a much longer time of exposure to external factors on the human thermoregulation system. For a small group of people, for example, athletes, it is possible to build statistical models limited to days, since most of the processes in the body are periodic with different cycles, mainly for general biology this is a day in terms of human heat production. In a long period of time in general biology, the adaptive mechanisms of the human thermoregulation system to the environment are studied. The temperature of the human "core" is 37 °C and is constant. Under comfortable conditions, the temperature on the surface of human skin is about 32 °C, although this value is different for the indigenous people of different climatic zones. At temperatures below 32 °C, signals on thermoreceptors begin to go on the skin surface and various biological and behavioral responses to the (effective) temperature of the environment in the human thermoregulation system arise. If the temperature exceeds 37 °C, diseases can occur in the internal organs. If a person is in uncomfortable conditions for more than 1 day, then in some cases this system can be considered in terms of general biology, for example, workers in the context of labor protection or athletes of the chosen discipline in the context of changes in their body for a long time. The characteristic size of an organ or body part is approximately 3-100 cm. Modeling a part of the body (arm, leg) of a person in the tasks of designing clothes or modeling an internal organ for surgery (cryo-, radiology, thermal, ultrasound, or other) to optimize external exposure and contraction the time of the procedure is performed with a characteristic size of up to 1m in computer programs. Usually, the procedure or surgery takes about 5-30 minutes. The time order for the 12 o'clock body part is chosen as work shift or service time. If the external heat / cold effect is aimed at irreversible destruction of tissue for the organs of a living person, then its temperature goes beyond the range of 0-45 °C, since when exceeded, irreversible denaturation of the protein (many proteins) occurs, and below it, ice is formed that can disrupt the membrane cells. Any biological tissue consists of cells, the characteristic size of which is 0.005 mm, although there are cells of different sizes even in the blood. The destruction of cells is facilitated not only by temperature, but also by its rate of supply or removal from the cryodestructor or thermode. There are other factors, within the framework of the use of high and low temperatures in medicine, for the destruction of pathological formations, for example, microwave radiation during cryo-operations. All these procedures of thermal exposure to diseased biological tissues last no more than half an hour on average per session. On average, within about 30 seconds from a short-term exposure to high or low temperatures, a danger signal is generated by neurons and measures are taken to minimize this temperature effect on the body due to the thermoregulation system - its various mechanisms (behavioral, vascular reactions, others). At the cellular level, at temperatures below 34 °C and above 38 °C, the process of redox reactions begins, aimed at maintaining the body's homeostasis, and energy is also released during the breakdown of ATP in the form of heat. The cellular level of the description of temperature processes is considered in the context of a living person, and not in vitro test-tube studies of cells. When the cells are cooled to the formation of ice, the sodium pump is disrupted and the cell swells; the phase state of lipids of cell membranes changes and the activity of membrane enzymes is disrupted, the pH of solutions inside the cell and in the intercellular fluid changes. The characteristic time of protein synthesis is about 1 second, and in 30 seconds it is possible to describe an ensemble of cells (in fact, biological tissue) with macroeffects, for example, a wavy (not Fourier) flow of heat waves. With a characteristic time of 30 seconds or more, biosystems are considered within the framework of molecular biology, and it is better to describe these problems by methods of general cell biology, for example, in the search for aging mechanisms. This has little to do with thermophysical aspects and problems and is not considered in the context here. At the molecular level of the description of thermophysical aspects, the characteristic size of a water molecule is $10^{-10}$ m and the vibration time of bonds in biostructures is $10^{-12}$ s, where the thermal vibration of molecules with a characteristic time of $10^{-9}$-$10^{-12}$ seconds is taken into account. Regardless of the type of nucleation (homogeneous or heterogeneous), crystallization centers begin to appear at a temperature of about 0 °C. At a temperature of about 45 ° C, proteins unfold with a decrease in the order of their structure (some from 4 to 3 and from 3 to 2), amino acids lose their tails, and under the action of an increasing hydrogen bond due to the release of "bound water", the main chain links begin to break. This is facilitated by short-term exposure to high temperatures, the decomposition products of amino acids and proteins, and at a time of the order of several seconds at low temperatures, the humoral and hormonal systems of thermoregulation act, which, through neural connections, trigger the synthesis of proteins and other biostructures or "uncouplers" (such as thermogenin) for the breakdown of ATP.

Using the methods of elementary combinatorics, it is possible from Table 1 to obtain the System F (S, L, t, T) = Fi, i = 1 ... 21 within the framework of the criterion of scale, time and temperature for systematizing tasks and aspects of temperature effects on humans. In this case, the rule must be observed that the macro level of the description of a lower system in terms of the size of the biosystem (S) must reflect the features of the description of a higher level system. That is, when describing at the macro scale (S5), must satisfy the description of the system (S4), and so on. The criteria under which a biosystem from one level should be described and modeled by methods from another biosystem are also taken into account. For example, if the size of biological tissue is more than 100 mm, then it is better to describe and consider it as an organ. If the time of thermal effect on biological tissue is less than 0.5 minutes, then due to transient processes it is difficult to judge its state after a short-term interaction and it is more logical to describe and consider this system at the cellular level in order to better understand the processes. To go from one level of the description of a biosystem (S) to another, it is enough that only one of the characteristic parameters (L, t, T) goes beyond the indicated framework. This is not a strict sentence, but a logical step in creating a future expert analytical computer system for individual calculation of the body's thermophysical reactions to external influences in understanding the human thermoregulation system and other local thermophysical aspects in planning and treatment or recovery from illness, as well as for labor protection tasks ... It should also be borne in mind that there are biorhythms of the whole organism, organs and cells of which it consists. These rhythms are developed by the evolution and the daily rhythm of the planet. Biorhythms are especially important in studies of human physiology for deep space flights.

From Table 1, the classified groups of tasks are obtained:

1. A group of people is exposed to cool and low temperatures for a short time up to 1 day, periodic studies are possible for up to 1 month. $F(S1, L2, t2, T1)=F1$
2. A group of people is in a comfortable temperature for a short time up to 1 day, periodic examinations are possible for up to 1 month. $F(S1, L2, t2, T2) = F2$ - this does not apply to thermophysical studies.
3. A group of people is exposed to high temperatures for a short time up to 1 day, periodic studies are possible for up to 1 month. $F(S1, L2, t2, T3)=F3$
4. A group of people is exposed to cool and low temperatures for a long time more than 1 day, periodic studies for up to 1 year or studying the dynamics of adaptation of a generation or more are possible. $F(S1, L2, t3, T1)=F4$
5. A group of people is in a comfortable temperature for a long time more than 1 day, periodic studies are possible for up to 1 year or the study of the dynamics of adaptation for a generation or more. $F(S1, L2, t3, T2) = F5$ - this does not apply to thermophysical studies.
6. A group of people is exposed to high temperatures for a long time more than 1 day, periodic studies for up to 1 year or the study of the dynamics of adaptation for a generation or more are possible. $F(S1, L2, t3, T3)=F6$
7. A person is exposed to cool and low temperatures for a short time up to 5 minutes, periodic examinations are possible for up to 1 month. $F(S2, L2, t1, T1)=F7$
8. A person is in a comfortable temperature for a short time up to 5 minutes, periodic examinations are possible for up to 1 month. $F(S2, L2, t1, T2) = F8$ - this does not apply to thermophysical studies.
9. A person is exposed to high temperatures for a short time up to 5 minutes, periodic examinations are possible for up to 1 month. $F(S2, L2, t1, T3)=F9$
10. A person is exposed to cool and low temperatures for up to 1 day, periodic studies are possible for up to 1 month. $F(S2, L2, t2, T1)=F10$
11. A person is in a comfortable temperature for up to 1 day, periodic examinations are possible for up to 1 month. $F(S2, L2, t2, T2)=F11$
12. A person is exposed to high temperatures for up to 1 day, periodic studies are possible for up to 1 month. $F(S2, L2, t2, T3)=F12$
13. A part of the human body (or organ) is at a temperature of 0-45 ° C for up to 12 days, periodic studies are possible for up to 1 month. $F(S3, L2, t2, T2)=F13$
14. Biotissue is subject to destruction by low temperature within 0.5-30 minutes, several cycles of freezing and thawing are possible. $F(S4, L2, t2, T1)=F14$
15. Biotissue is in a living state at a temperature of 0-45 ° C for 0.5-30 minutes. $F(S4, L2, t2, T2)=F15$
16. Biotissue is subject to destruction by high temperature within 0.5-30 minutes, several heating cycles are possible. $F(S4, L2, t2, T3)=F16$
17. The cell is exposed to low temperatures for 1-30 seconds, several cycles of freezing and thawing or reversible processes are possible. $F(S5, L2, t2, T1)=F17$
18. The cell is at a temperature of 34-38 ° C for 1-30 seconds, the processes in the cell are reversible and it is alive. $F(S5, L2, t2, T2)=F18$
19. The molecule is exposed to low temperatures for 10 -12-1 seconds. The balance of retention of the chain of molecules is disturbed due to the increase in the influence of hydrogen bonds. $F(S6, L2, t2, T1)=F19$
20. The molecule is part of a liquid medium at a temperature of 0-45 ° C for 10 -12-1 seconds. Hydrogen bonds in the chain of molecules are balanced by van der Waals forces. $F(S6, L2, t2, T2)= F20$

21. The molecule is exposed to high temperatures for 10 -12-1 seconds. The balance of retention of the chain of molecules is disturbed due to the increase in the influence of hydrogen bonds. F(S6, L2, t2, T3)= F21

Of the 21 options considered, which did not include quantum processes, 18 options belong to the field of thermal physics. When describing the state of any complex system, a group of external factors and influences and a group of internal processes and relationships are distinguished. According to this principle, Table 2 can be drawn up. Understanding all the factors influencing a person will allow us to determine his current state and assess the future state of the body. This will allow you to diagnose diseases and the state of the body before treatment.

Table 2. Main external and internal factors for different biosystems

|   | Biosystem size (S) | External factors | Internal factors |
|---|---|---|---|
| 1 | Group of people (S1) | 1) environmental impact (physical)<br>2) behavioral (lifestyle) | 1) diseases<br>2) age<br>3) adaptive mechanisms of survival and mutation |
| 2 | Human body (S2) | 1) environmental impact (physical)<br>2) behavioral (lifestyle)<br>3) food composition | 1) diseases<br>2) age<br>3) Gender (sex)<br>4) individual characteristics of the body (weight, physical fitness, features of organs and their functioning, etc.) |
| 3 | Body part (organ) (S3) | 1) environmental impact (physical)<br>2) the presence of foreign bodies (prostheses, biochips) | 1) diseases<br>2) age<br>3) individual characteristics of the body (weight, physical fitness, features of organs and their functioning, etc.) |
| 4 | Fabric (S4) | 1) environmental impact (physical)<br>2) the presence of foreign bodies (prostheses, biochips)<br>3) artificial influence (physical) | 1) diseases<br>2) individual characteristics of the body (weight, physical fitness, features of organs and their functioning, etc.)<br>3) functioning of the circulatory system in and near tissue<br>4) functioning of the nervous system |
| 5 | Cage (S5) | 1) environmental impact (physical)<br>2) artificial influence (physical) | 1) diseases<br>2) age<br>3) rate of biochemical reactions<br>4) membrane properties<br>5) the ability of nuclei in cells |
| 6 | Molecule (S6) | 1) artificial influence (physical)<br>2) natural impact (physical) | 1) rate of biochemical reactions<br>2) restructuring rate of macromolecules<br>3) presence of viruses |

From the point of view of thermal physics, the classification presented in Table 1 according to the level of description of biosystems (S) can be divided into groups of specific tasks and problems not according to the classification system F (S, L, t, T), but according to the principle of division of sciences (C), highlighting basic: general biology (C1), physiology (C2), medical physics (C3), biophysics and biochemistry (C4). Each scientific approach (C) can be divided into groups of tasks. And then it is more convenient to represent this division in the form: general biology (C (N1, j)), physiology (C (N2, k)), medical physics (C (N3, l)), biophysics and biochemistry (C (N4, m)). In each section of science, one can single out the main tasks related to the study of thermoregulation and thermophysical aspects in the human body. And also to highlight other sciences and directions in them, but this will only complicate the narration and classification of this work in the understanding of multidimensionality. It is proposed to indicate only the main directions of studying thermoregulation in the classification C (N (1, j)), C (N (2, k)), C (N (3, l)), C (N (4, m)) for visualization directions of research of thermoregulation in science.

For general biology C (N (1, l)), j = 1,2,3 ... for problems of studying a group of people (S1):

1. Climatic zone.
2. Profession (active and passive activity).
3. Acclimatization.
4. Weightlessness.

For physiology C (N (2, k)), k = 1,2,3 ... for tasks of studying the human body (S2):

1. Body systems (respiratory, circulatory, nervous and others).
2. Gender, age.
3. Thermal regulation mechanisms in the nervous system.
4. Physical mechanisms of heat release.
5. The reaction of the bloodstream to temperature changes.

For medical physics (C (N3, l)))), l = 1,2,3 ... for tasks on the study of organs (S3) and tissues (S4):

1. Physical 3D computer models of organs and body parts.
2. Physical properties of biological tissues.
3. Bioheat and mechanisms of heat transfer in biological tissues.
4. The rate of freezing and thawing of biological tissue.
5. Thermal burns. Tissue necrosis from cold.

For biophysics and biochemistry (C (N4, m)))), m = 1,2,3 ... for the tasks of studying cells (S5) and molecules (S6):

1. Diseases are different (classification by ICD is possible).
2. Composition of cells and intercellular fluid.
3. Oxidative mechanisms, conduction of nerve channels, activators of chemical reactions, membrane potentials, reaction rates.
4. Individual sensitivity (psychophysics or sensing, genetics).
5. Phase processes, osmosis, nucleation, quantum processes.

It is possible to further refine the classification of C(N (1, j)), C(N (2, k)), C(N (3, l)), C(N (4, m)) and possible tasks that can be attributed immediately into two or more subgroups in such a classification, or develop more precise criteria for describing the subclasses N(1, j), N(2, k), N(3, l) and N(4, m).

**2.2 The solution of the problem.**

**In this paper, it is proposed to superimpose the classification model C(N (1, j), C(N (2, k)), C(N (3, l)), C(N (4, m)) on the classification model F(S, L, t, T) in the form of clarifying approaches to the study of thermophysical aspects and the system of thermoregulation in humans.**

It is worth noting that the **combined F[C] model** has dead-end areas of common sense research. For example, they do not study a person's prolonged stay in extreme temperatures that are incompatible with life (even in the field of disaster medicine, the percentage of affected skin, the degree of burns is established, and the question of the time of the remaining life is not studied, according to the "Hippocratic Oath" and the "Nuremberg Code").

A review of the most significant scientific literature can be performed by completing Table 3 "Generalized model for the classification of thermoregulation and thermophysical aspects for humans" for the combined model **F[C]**.

Table 3. "Generalized model of classification of human thermoregulation" for the combined model **F[C]**.

|  | N(1,j) | j=1,2,3… | N(2,k) | k=1,2,3… | N(3,l) | l=1,2,3… | N(4,m) | m=1,2,3… |
|---|---|---|---|---|---|---|---|---|
| F(S1, L2, t2, T1)=F1 | [], [], [] (review article) |  |  |  |  |  |  |  |
| … | [], [], [] |  | [], [], [] |  | [], [], [] |  | [], [], [] |  |
| F(S6, L2, t2, T3)= F21 |  |  |  |  |  |  | [], [], [] (review article) |  |

The purpose of this work is not to fill in Table 3 "Generalized model for the classification of human thermoregulation" with all scientific articles. Any scientific direction develops iteratively - from simple models to complex ones, complex models are combined into systems, systems into general multicriteria nonlinear problems that are solved using computer technology and artificial intelligence. Table 3 serves as a way to systematize the classification of thermophysical problems for humans.

To compose a more complex system of equations describing human thermoregulation, it is necessary to have a large amount of experimental data. For this, it is necessary to improve the measuring technique and measurement methods to study various physical properties of the skin and the general condition of the body. Thanks to the systematization of knowledge, it is possible to identify unexplored areas and unite specialists from different sciences to solve the identified urgent problems in interdisciplinary modern science. In this case, the mathematical rules for physical modeling of natural phenomena (processes) on models can be taken from [205].

**2.2.1 Factors according to the classification F [N1] (group of people: general biology).**

Attempts to study the impact of a changing climate on the planet have pushed researchers and engineers in recent decades to study the impact of climate on humans. This has found practical application in labor protection tasks with a description of the conditions for a comfortable state of a person. Committees from different structures of one country and a number of countries were created to study the impact of Earth's climate change on humans. The engineering approach suggests calculating the comfort zone for a person using an online calculator with the issuance of results how the body will behave in response to weather changes, the forecast of which is carried out by the meteorological service. This could predict special measures in enterprises to protect workers from overheating or hypothermia. Understanding the response mechanisms in the human body to changes in the environment and the automation of calculations of effective environmental indicators based on the data of the meteorological service requires further development.

An important study of the comfort zone [23] was carried out in China, which shows a significant difference in the systems of thermoregulation of people living in different climatic zones.

In [196], the mechanisms of human acclimatization to a cold environment are described in detail, comparisons are made by sex, age, and reference literature is given.

Classification F [C], proposed in the work, shows the need to take into account in detail all external environmental factors and detailed information about the internal mechanisms of the human body. The most interesting tasks in the F [N1] direction are shown in Table 4.

Table 4. F [N1] factors

|   | F [N1] factors | Literature |
|---|---|---|
| 1 | Influence of the nature of work (stress) | 143 |
| 2 | Influence of climatic zones of residence on human physiology | 140, 141 |
| 3 | The speed and degree of adaptation to a new climate in humans | 138, 139, 142 |
| 4 | Interspecies differences | 10, 11 |
| 5 | Effect of aging on heat transfer in the human body | 213, 214 |

**2.2.2 Factors according to the classification F [N2] (human physiology).**

From the general course of physiology of human heat production [2 or 232], the following is known about the thermoregulation system. For the normal functioning of the human body, it is necessary to maintain homeostasis - a constant body temperature. Allocate the "core" of the body, which the central nervous system (CNS) maintains at a constant temperature due to the mechanisms of thermoregulation. The central nervous system receives signals from cold and heat receptors, which are located in the skin in different layers, as well as heat receptors located in internal organs and even blood vessels. The main part of the brain responsible for thermoregulation is the hypothalamus (the central link in thermoregulation), although impulses with a signal about temperature changes come to other parts of the brain and to the cerebral cortex. The main outgoing signals from the hypothalamus regulate heat transfer and heat production in the human body, but there are other parts of the brain and cerebral cortex, from which certain signals can emanate for the formation of human thermoregulation mechanisms. Thus, it was found that the parts of the spinal cord give signals for muscle contraction (cold muscle tremor - "tremor"). The hypothalamus itself, in terms of physiology, is divided into different zones that are responsible for different mechanisms of thermoregulation in the body. The posterior hypothalamus is the center of heat production and, with the help of the sympathetic nervous system (an efferent link of thermoregulation), increases the tone of the body by increasing catabolism. The anterior hypothalamus is the center of heat transfer and acts with the help of the parasympathetic nervous system on the heart rate, reducing its work due to anabolism. The central nervous system sends specific signals for the three main mechanisms of human thermoregulation, which in turn are divided into subsystems.

1. Conscious human actions determine his **behavioral system of thermoregulation**.
2. **Mechanism with vascular reactions.** Between the arteriole and the venule there is a regulated blood vessel - astamosis (arteriovenous shunt), located close to the capillary bed. With an external cold environment, there is no blood in the capillaries, it passes through astamosis, the diameter of which at this time is increased in comparison with a comfortable state.
3. **Hormonal and humoral systems of thermoregulation.** Humoral thermoregulation is carried out at the expense of the endocrine glands (mainly thyroid and adrenal glands), whose hormone production is regulated by the nervous system. Thyroid hormones increase the body's metabolism. The adrenal glands produce adrenaline, which enhances oxidative processes in tissues, increases heat production and constricts blood vessels, reducing heat transfer.

3.1. <u>Heat production mechanisms</u>.
    3.1.1. Contractile thermogenesis.
        3.1.1.1. Voluntary muscle contraction.
        3.1.1.2. Thermoregulatory tone. Muscle tone is increased due to the extrapyramidal system. Increases total heat production by 20-40%.
        3.1.1.3. Cold muscle shivering. A more intense increase in heat production, since non-synchronous muscle shivering release energy, which is mainly heat.
    3.1.2. Non-contractile thermogenesis.
        3.1.2.1. Changes in the intensity of general cell metabolism.
        3.1.2.2. Increased heat production from internal organs through which blood flows. This is mainly the liver.
        3.1.2.3. Dissociation of redox reactions (RR). In mitochondria, ATP is usually formed in the electron transfer chain. Under the influence of various "uncouplers" (such as the protein thermogenin), ATP breaks down and the released energy turns into heat.
        3.1.2.4. When decaying, brown fat releases a lot of energy, which turns into heat. There is a lot of brown fat in babies, but in an adult it is almost nonexistent.
        3.1.2.5. Specific dynamic action of food. Energy consumption for the digestion of food products increases.

3.2. <u>Heat transfer mechanisms</u>.
    3.2.1. Physical mechanisms. Radiation (radiation transfer) heat transfer dilates blood vessels and heat transfer in the body increases; takes about 60% of the total heat transfer. Evaporation occurs at the expense of the sweat (and sebaceous) glands on the skin, which are controlled through sympathetic innervation (acetylcholine is released, which acts on the glands through the m-choline receptors); takes about 15-20% of the total heat transfer. When in contact with solid surfaces, heat is released due to thermal conductivity and is about 5-10% of the total heat transfer. In a liquid or gas, heat transfer occurs by a convective mechanism. These physical mechanisms work both for heat transfer and heat transfer to a person from hotter bodies.
    3.2.2. In the heat, the metabolic rate of the whole organism decreases.
    3.2.3. Loss of heat in feces and urine.

    The temperature of the "core" of the human body is also subject to change. Hypothermia has several meanings. Hypothermia as a disease (exogenous nature). When a person is in a hot environment for a long time, where his thermoregulation system could not withstand the load and the temperature of the "core" rises. It is treated with medication. Hypothermia as a controlled process. The doctor lowers the temperature of internal organs in order to increase the resistance of tissues to hypoxia, that is, so that the tissues would live longer in the absence of oxygen (they are used in organ transplantation). Temperature control of body tissues is done using ice or the administration of muscle relaxants to reduce contractile thermogenesis, or substances (such as chlorpromazine) that act on the thermoregulatory center (hypothalamus). Hypothermia as a disease, fever (endogenous nature). With fever, the hypothalamus perceives signals about the external temperature with a decrease in their actual values. The body triggers the mechanisms of heat generation and thus a person has a fever, and he feels cold. Fever occurs due to pyrogenic substances: endogenous pyrogens (such as cytokines) affect the factors of the immune system; exogenous pyrogens (such as lipopolysaccharides, components of bacteria) affect the overall metabolism of cells. Pyrogens eventually shift the set point in the hypothalamus from the normal temperature perception by receptors.

The metabolic rate determines the heat production of the body and depends on a number of factors. The individual characteristics of the organism, influencing metabolism, related to its main general parameters - gender, age, weight, height, body surface area and the ratio of mass to surface area - were revealed. In newborns, heat transfer is higher than in adults due to intense blood circulation and evaporation from the skin surface. With age, the thermoregulation of an adult acquires the ability for contractile thermogenesis, reflex mechanisms are developed as a response to environmental changes. In old age, the human thermoregulation system regresses. [3] At the same time, with a decrease in the ambient temperature, thermogenesis increases, as with intensive muscle work. The nature of the diet (quality and quantity) affects thermogenesis. The emotional state of a person increases heat production and a person tolerates greater cold. Hypoxia lowers the thermogenic threshold of the hypothalamus and thermal sensitivity. [173] The intensity of heat production decreases in the dark and under intense ultraviolet irradiation. [2]

In the online lecture of Professor Hanaa, in tabular form in English, the human thermoregulation system is available and briefly described from a physiological point of view. [12]

In the ScienceDirect system, at the request of "Metabolic Heat Production", a page is formed with a selection of information on heat production with abstracts of articles (https://www.sciencedirect.com/topics/engineering/metabolic-heat-production), where such works as [6] and a number of works to determine thermal comfort and the effect of clothing on human heat production. The review work [6] investigates thermoeffective reactions during heat stress, as well as the mechanisms of thermoregulation. Aspects related to non-thermal modulators of thermoeffective reactions are explored and the effects of body composition, aerobic fitness, heat acclimatization, gender, age, chronic diseases (eg diabetes), hydration and cardiovascular function on the body's ability to dissipate heat are discussed.

In the PubMed database, you can search for similar articles by the identifier of a specific article in the PMID system. The query "comfort" contains many articles on OSH that provide different empirical indicators (HSI, ET, CET, ET *, P4SR, WBGT, UTCI). Of these models, it is worth highlighting the UTCI thermo-climatic indicator, created by eight organizations from different countries, as the most generalizing in this area of research and the most universal temperature indicator. The UTCI mathematical model predicted a thermoregulatory response in accordance with thermal equilibrium conditions under different environmental conditions. This model defines a universal indicator of climatic conditions, since the air temperature in "reference", baseline conditions (simulated) produces the same response as real conditions. The UTCI mathematical model does not take into account environmental conditions, but the physiological response to them. An online UTCI calculator was created (http://www.utci.org) [19]. The work [20] compared the universal temperature climatic index (UTCI) with the selected thermal indices / environmental parameters for 12 months a year.

In recent years, scientific research has been devoted to climate change and its impact on humans. Review work [113] is devoted to this topic in the context of global warming.

The tasks for section F [N2] physiology have already been sufficiently studied and described in the scientific literature from different countries, but they have not been combined into a single base. The creation of such a base is a promising task for creating a personalized thermoregulation system and other areas in the future. The most interesting quantitatively dependences for understanding the human thermoregulation system according to the F [N2] classification can be distinguished in Table 5.

Таблица 5. F[N2] факторы

| | F [N2] factors | Literature |
|---|---|---|
| 1 | The presence of hair on the body, its unevenness, polymotor reflex | 112, 114, 115 |

| 2 | Atmosphere and humidity | 119, 120, 121, 122, 123, 124 |
|---|---|---|
| 3 | Assessment of heat generation from hypodermis at low temperatures in different anatomical regions of the human body | 125, 126, 127 |
| 4 | Assessment of heat generation from the work of different muscles of the human body | 128, 129 |
| 5 | Fluctuations in human body temperature during the day or more | 130 |
| 6 | Dependence of the thermal emissivity on skin color | 60 |
| 7 | Study of the influence of extreme factors on changes in the physiological functions of the body | 174, 175, 178, 193 |
| 8 | Mathematical relationship of the thermoregulation system with other systems in the body | 6, 131, 132, 134 |
| 9 | Taking into account the acid-base properties of skin and blood. Accounting for lipid composition. | 109, 110 |
| 10 | Heat production of human organs and muscle energy. | 7, 8, 9, 144, 145 |
| 11 | The effect of sleep on the state of heat production | 25, 117 |
| 12 | Assessment of local effects on the state of comfort on an individual basis | 26 |
| 13 | Transitional blood temperature from the ambient temperature gradient | 40 |
| 14 | The influence of the genotype on receptors and more | 86 |
| 15 | Influence of age | 116, 118 |
| 16 | Respiratory intensity and work patterns (mental and physical) at high and low temperatures | 135, 136 |
| 17 | Age-related changes in the thermoregulation system | 215, 216, 217, 218 |

Local and general thermophysiological models were created for humans, which eventually became multi-segment models in computer modeling. The following thermophysiological models of a standard human body are best known. The 1970 Stolwijk model [146] divided the human body into 5 cylindrical segments, each of which was divided into four concentric layers connected by blood flow. The 1999 Fiala model [147] divided the human body into 15 spherical or cylindrical elements. In 1971, the two-node models Gagge [148] and Givoni and Goldman [149] were proposed, representing the human body as a cylinder with a core. In 1977 Azer and Hsu proposed a two-node model [150]. In 1992, Jones and Ogawa proposed a two-node transient model [151]. In subsequent multi-layered models, the human body was divided into 5 layers: skin, fat, muscle and core or bone. Layered models: Wissler 1985 [152]; Ring and Dear 1990 [153]; Smith 1991 [154]; Fu 1995 [155]; 1968 Wyndham and Atkins [156]; Berkeley Comfort Model [157, 158] 2001; Huizenga et al. 2001 [159]; Tanabe 2002 [160]; Salloum et al. 2005 [161]; Al-Othmani et al. 2008 [162]; Thermo SEM 2012 [163], JOS-2 2013 [164]. Thermal Models Simulating Human Body Parts: Worrall 2012 [165]; Deshpande 2007 [166]; Ferreira 2012 [167]; Shitzer 1997 [168]. There appeared polysegmental models of the human body, created by a team of scientists: a voxel model in 2009 [132, 219] and a multisegmental model in 2018 [4]. A multisegmental model taking into account the circulatory system was compiled in 2015 [187], anastomoses accounting (AVA) was made in the model [220] in 2014. The newest neurophysiological human thermal model based on the reactions of thermoreceptors was compiled in 2020 [17]. More specific review articles are published in scientific journals, mainly on thermal comfort [5].

Modern work in the study of «Heat exchange between the human body and the environment: A comprehensive, multi-scale numerical simulation»[201]. More details about the thermoregulation system of the human body are written at the level of physiology and neurophysiology in the book [232].

**2.2.3 Factors according to the classification F [N3] (medical physics for tissues and organs).**

For a more detailed study of heat and mass transfer processes and aspects in the human body, it is possible to single out a number of tasks from the description of the thermoregulation system according to C (N2) (physiological) and consider them in more detail. The C (N3) classification will include a number of tasks from medical practice and labor protection.

Within the framework of medical physics, bioengineering and other interdisciplinary approaches in the study of the state of the body, its reaction to external and internal factors, physical and mathematical models are created for the purposes of medicine and labor protection. 3D models of human organs with different morphologies appeared, made in different software environments for training students, doctors, as well as for planning and optimizing treatment techniques (mainly surgical intervention). From the point of view of thermophysics, it is important to know the structure of biological tissue at the tissue level of the description of the system, since the propagation of heat can proceed along a parabolic (Fourier) dependence and along a hyperbolic one - by waves (not Fourier). In porous media (liver, lungs), heat and mass transfer has its own characteristics. Within the framework of local problems in thermomechanics, problems of the speed of the freezing front, its temperature, as well as the distribution of the heat front under thermal effects of different nature (hot metal objects, laser, medical accelerator) are often solved. Within the framework of labor protection, one can distinguish thermal phenomena that occur for a short time (up to about 5 minutes), for a long time and sometimes a periodic effect on parts of the human body.

In local problems of thermophysics of organs and tissues, it is important to take into account the individual physical characteristics of biomatter; its metabolism (cell division); heat production from muscle and fat; different responses of the body in the system of thermoregulation according to C (N2).

The main characteristic studies for F [N3] are shown in Table 6.

Table 6. F [N3] factors

|   | F [N3] factors | Literature |
|---|---|---|
| 1 | Influence of external (atmospheric and artificial) pressure | 52 |
| 2 | Influence of moisture and perspiration | 78, 116 |
| 3 | Thermomechanical and mechanical parameters | 44, 48, 49, 50, 63, 64, 65, 66, 67, 68, 69, 76, 77, 104, 106 |
| 4 | Taking into account the n-dimensionality of mathematical models | 101, 102 |
| 5 | Assessment of the influence of constant and alternating electromagnetic fields | 84, 85, 94, 95, 96, 183 |
| 6 | Accounting for ionizing and non-ionizing radiation of different power and secondary radiation, as well as laser heating | 51, 98, 105, 108, 134 |
| 7 | Anisotropy of physical properties and their dependence on temperature and other | 54, 55, 56, 34 |
| 8 | Correct calculation of the size of the measuring cell for blood perfusion, taking into account the temperature of venous and arterial blood | 70 |
| 9 | Accounting for turbulent air / liquid flows for the heat transfer coefficient between the body and air / liquid | 97, 60 |
| 10 | Complication of polysegmental general models of the human body towards fragmentation into smaller parts for the analysis and prediction of problems of modeling the thermal response to changes in external and internal factors | 28, 27, 4, 185, 186, 187 |
| 11 | Consideration of the location of the bloodstream | 29, 30, 31, 59, 204 |
| 12 | Complications of receptor response models due to their uneven location on the body, even at different depths of one part of the body | 32, 62 |

| 13 | Studying the characteristics of signals from thermo- and cold receptors | 32, 33, 57, 103 |
|---|---|---|
| 14 | Study of the rate of temperature change in tissue during cryodestruction | 34, 35, 36 |
| 15 | Physical foundations of cryobiology | 37, 39 |
| 16 | Phase transitions at low temperatures | 38, 41, 42, 43, 46, 47, 48 |
| 17 | Vascular tree models | 45 |
| 18 | Radiative heat transfer coefficients, as well as natural and forced convection coefficients. | 60 |
| 19 | Taking into account the effect of clothing on the heat balance | 61, 179, 180, 181, 182 |
| 20 | Thermal burns | 78, 80, 81, 82, 83, 89, 90, 91, 92, 100, 107 |
| 21 | Contact resistance between body and solid object | 198 |

It is especially worth noting a number of works of a general nature, which collect modern ideas about a particular thermophysical aspect or features of thermoregulation in the human body. Below are these works with their brief description.

The review "Thermophysiological models and their applications: a review" [5] presents the historical development of thermophysiological models of a local area of the skin, analyzes and compares the models with each other. Investigations of thermophysiological models of isolated body parts are presented. Several applications of the models are given, the most important issues for discussion are presented in detail. Review papers have been published earlier - "Thermal comfort models: a review and a numerical study" [24]. In further studies [4], the Pennes equation was solved for 16 body parts, taking into account the corresponding physiological properties for each segment. The Pennes equation control system was added using the thermoregulation mechanisms of the 65-node Tanabe (65MN) model.

In [58], a review of bioheat models for 2013 is given, which take into account various external and internal factors that are present in the heat exchange of human skin with the environment.

It is important to take into account the mechanical properties of biological tissues during thermal phenomena. This is the subject of work on the mechanical and thermomechanical properties of biomedical objects. The work [68] gives an overview of thermomechanical models of biological tissues for 2009, as well as in the work of 2018 "Mechanical properties and modeling of the skin: a review. Proceedings of the Institute of Mechanical Engineers "[75]. This knowledge is necessary not only in traditional areas such as cosmetology, but also in new ones - the "man-machine" environment, when sensors are implanted, internal organs are replaced, and neurointerfaces are connected [71, 72].

The effect of cryogenic temperatures on a medical and biological object is considered in the work of A. I. Zhmakin "Physical foundations of cryobiology" [37], which describes physical phenomena at the molecular, cellular levels, at the level of tissues and organs in biological objects during freezing and thawing. Optimal values of the temperature of the cooling gaseous heat carrier and the duration of its interaction with the surface of the object of cooling (human body) have been found for devices of general cryotherapy effect [53]. A review of cryotherapy techniques and methods of their exposure to humans is given in the review [176].

The work [82] describes in detail the mechanisms of the onset of pain from exposure to high temperatures. Thermal burn modeling is usually calculated using the Arrhenius model, no analytical three-dimensional solutions have been found, and the numerical solutions and models are different, similar to the solutions of heat conduction problems without phase transformations. Another approach to predicting a thermal burn for a local part of the human body is given in [89], where the problem was solved numerically by the method of a matrix of heat transfer lines. The heat balance equation is written as the Pennes equation, and the rate of irreversible protein denaturation (thermal burn) was calculated using the Arrhenius model. A two-dimensional matrix model of the transmission line was used to predict the effects of the impact.

Critically analyzed a review of the evidence linking temperature and time to cell death and the depth of burn injury. This is necessary, according to the author, for a better understanding of information on time-to-temperature relationships across a wide range of industry standards, burn prevention literature, and medical judgment. [93]

In [102], a detailed analysis of the Pennes equation (bioheat) is given, and the study of temperature dependences on various parameters in the modern formulation of the bioheat equation is given: on the influence of blood flow, a different nature of external influence, on various kinds of boundary conditions. The problem is solved in the Cartesian coordinate system in space (3D). The main value of this work, according to its authors, is the solution of the bioheat equation for various kinds of external influences, which are most often encountered from practical examples in various fields of medicine.

In medicine, the laser is increasingly used for treatment. Models were created for heating the skin from such radiation. On the one hand, this is a thermomechanical task, and on the other hand, it is a special external influence. Such problems are solved numerically and the solution is visualized by a graph for a local area of human skin, as, for example, in [105].

Different types of sensors are used to locally study human skin. A recent review of the most commonly used sensors for studying the properties of the skin is given in [99].

Among the latest works, it is worth noting the doctoral dissertation of 2019 "Mechanical and biochemical response of human skin to various load conditions." [74]

**2.2.4 Factors according to the classification F [N4] (biophysics and biochemistry of cells and molecules).**

The mechanisms of heat generation in the body are more associated with physicochemical processes in cells, in the intercellular fluid and mechanisms at the molecular level. Thanks to signals from the central nervous system through the nervous system, the mechanisms of humoral and hormonal heat production and heat maintenance in the body are triggered due to the synthesis of enzymes, proteins, amino acids. And those, in turn, due to the reactions that are studied by the methods of molecular dynamics and quantum physics, affect the functioning of organs, the abilities of tissues, cell metabolism and other mechanisms, releasing energy in the form of heat.

Molecular dynamics (MD) methods *per se* and others are now the main tools for studying biophysical problems. They model molecules of proteins, nucleic acids, study nucleation, the growth of ice crystals and its shape, the processes of melting and thermal denaturation of proteins, osmosis. Different potentials on cell membranes and potentials of water in subcellular structures are simulated by computer.

MD methods are often needed to study not only the mechanism of physicochemical at the micro level, but also to calculate the coefficients necessary for the macroscopic description of heat and mass transfer in biological systems at the level of cells and tissues. For this, macromolecules are described as flexible chains. Hybrid methods are being developed for describing a group of molecules and the bonds between them at once [189], which makes it possible to simulate more complex effects. In practical terms, this gives, for example, the choice of the most suitable cryoprotectant (solution) for cryo-operations.

A review of works on modeling bilipid membranes at room temperature by MD methods performed before 2000 is given in [190], another review of lipid membranes is presented in 2015 [191]. Much research has been done to study ion channels and their collective dynamics for physical modeling of transport in biological structures. This is important when creating cryoprotectants (injections and ointments) for cryo-operations on various human organs.

When describing biological systems, different models are used that differ in the characteristic size and time of the modeled process. There are not always operators of the transition from smaller and short-term models to larger models with their fluctuations at the macroscopic level and effects at the macroscopic level. For example, simulation models of ice propagation at the cellular level.

The most difficult task is to describe the mechanisms of influencing factors according to the classification F [N4]. Such factors include diseases, the effects of drugs, poisons, pregnancy, the presence of foreign bodies, the influence of bad habits, and more. There is no mathematical model to describe quantitatively the relationships between the systems of the body, when one of its systems begins to work in the mode of illness / recovery and illness. Physicomathematical problems in the biophysics of subcellular structures are complex in their mathematics, and their practical application has not yet been found for wide use in the field of describing the individual system of thermoregulation for humans, but is more used in dermatology and in the scientific and theoretical direction.

It is especially worth noting a number of works of a general nature, in which modern concepts of various biophysical and biochemical mechanisms at the level of cells and molecules that are involved in the processes of thermoregulation in the human body are collected. Below are these works with their brief description.

The mechanism of sensitivity to temperature at the molecular level is based on a change in the ionic conductivity of channels formed by special proteins. The spike frequency of neurons depends on the ionic conductance of the channel, which, in turn, depends on the temperature. In recent decades, several proteins from the TRP (transient receptor potential) family have been discovered, whose temperature response profiles cover the entire physiological temperature range. These proteins are considered the most likely candidates for the role of molecular temperature sensors [14]. However, it is assumed that there are other, as yet unknown molecular mechanisms of thermal sensitivity [15], and research in this area continues and new mechanisms of signaling to the brain from receptors are being discovered (a new phylogenetic pathway that transmits high-precision cutaneous thermal sensory activity in the form of "marked lines") [16]. A new neurophysiological human thermal model based on the reactions of thermoreceptors was compiled in 2020 [17].

Mechanoreceptors are reviewed in the work of A.V. Zeveke. "On the theory of skin sensitivity" [21], which shows the mechanisms of formation of codes for cold, pain, touch and heat only on the basis of skin mechanoreceptors. Further work of the author and his school can be found in publications, for example [22]. The rest of the work is a more detailed study of biophysical mechanisms at the subcellular level. It should be noted that there are different receptors in the skin and internal organs and they have different amplitude-phase characteristics in time upon activation. The activation mechanisms are different and different signal transmission pathways for each type of receptor with different frequency characteristics of signal transmission to the central nervous system. Not many experiments have been carried out in this area on humans to observe the evolution of receptors from age, habitat (climatic zone) or human genotype, and other factors.

It is also necessary to take into account the biophysical basis of pain sensations [87], which differ depending on the genetic type of a person [86]. Modeling the onset of pain from high temperature is shown in [88].

In work [37] "Physical foundations of cryobiology" a large review of models, including biophysical ones, is given. In work [73] "Mathematical and computational modeling of skin biophysics: a review. Proceedings of the Royal Society "provides biophysical foundations of modeling and basic concepts in the system of thermoregulation for human skin.

The paper [197] "Methods of molecular dynamics for modeling physical and biological processes" provides an overview of the current state of research in the field of computer modeling of physical and biological systems using molecular dynamics methods. The features of computer modeling of molecular and atomic systems based on parallel and vector systems are considered.

Basic textbooks in the study of biophysics of molecular and submolecular biological structures can be divided into 2 books, where the material is most fully and structurally presented - "Methods in Molecular Biophysics: Structure, Dynamics, Function" [199] and "Biophysics. Textbook in 2 volumes "[200]. There is a huge number of books and monographs on separate sections of biophysics for further more detailed study of the material. For example, computational cell biology is described in the book [206].

A better understanding of the processes inside the cell will require knowledge of biological chemistry. Modern books [207, 2008] are written in the form of encyclopedias, reflecting modern ideas in this area of knowledge.

The role of the brain (CNS) in the work of thermoregulation is discussed in the work "Central Mechanisms for Thermoregulation" [231].

The main characteristic studies for F [N4] are shown in Table 7.

Table 7. F [N4] factors

|   | F [N4] factors | Literature |
|---|---|---|
| 1 | Role of cytokines | 194 |
| 2 | Mechanism of hypothermia and pathophysiology at the cellular level | 195 |
| 3 | Heat dissipation by convection and evaporation | 221 |
| 4 | Biophysical Basics of Thermoregulation | 52, 222 |
| 5 | Acclimatization: Protein Reactions | 223, 227 |
| 6 | The effect of disease on thermoregulation | 224, 229 |
| 7 | Impact of pregnancy on thermoregulation | 225, 226 |
| 8 | Influence of hormones | 228 |
| 9 | Central mechanisms of thermoregulation | 231 |

**2.3 Problems and solution alternatives**.

The designated problem of studying complex biological systems of the human body, in particular its thermoregulation system, is now being solved by particular problems from various fields of science. Attempts are made to combine approaches in the study (active and passive systems of thermoregulation), use polysegmental models, describe in quantitative terms the reactions of human thermogenesis mechanisms depending on a set of environmental factors as one system with predictable responses of the body. If a complex system cannot be solved immediately, it is solved iteratively through special cases - this is the approach now. All special cases can be systematized, showing what else needs to be studied. On the basis of the collected experimental results, it is possible to construct a tree for predicting the state of the human thermoregulation system in a quantitative ratio as a multifactorial solution to the problem. In this work, a multifactor model, or a tree of tasks for predicting the state of a person, is proposed, but some research teams now look sharply negatively at such an approach in solving. Multivariate models are now being solved by special computing centers (using optimization criteria, taking into account the relationship between input data, intermediate patterns). The classical approach in the form of constructing ROC curves with an equation from variables is not applicable here as a method, as well as statistical methods of trees with different options for constructing them. Mathematical models for describing a complex living system (human) have not yet been created at the cellular and intracellular levels, which would make it possible to model the state of the body in diseases. But this does not mean that in the near future mathematical models will not be created for describing the organism as a whole (a person with its systems) at the tissue level as simulators for testing treatment tactics, diagnosing a disease or other application (military use - protection against chemical, biological, nuclear , information and psychological weapons and the creation of the soldiers of the future).

In order to obtain a practical result from the thermophysical aspects, it is necessary to be able to predict the state of the modeled system at different levels of its description. Forecasting is based on the following principles: dynamics of development of an organism, diseases, external factors of influence.

At the moment, each science builds its own physical and mathematical models to describe specific problems and effects that have their own limitations in modeling. Generalized principles and software systems have not yet been created to describe all the observed thermophysical aspects in the human body. In particular, the human thermoregulation system has not been fully described, and scientific discoveries in the field of biophysics of proteins, which are responsible for the synthesis of more complex bonds in the body, continue. In addition, the central nervous system has not yet been deciphered - where and which parts of the brain receive signals, how these thousands of signals of different amplitude-frequency characteristics are distinguished; how the brain "calculates" the signals that must be issued as an outgoing signal to control not only the thermoregulation system, but also other systems in the body. Thanks to these outgoing signals, certain compounds begin to be produced in the organs, which support the body in a multicriteria optimal state.

The classification of tasks is built in the form of packing some studies into others, thereby obtaining a multifactor predictive model of the thermal state of a person and his local part of the body. It is a statistical method for solving complex problems. There are semi-analytical methods for describing complex biological systems. A short review [204] provides a mechanism for generating simple models from complex ones (asymptotic decomposition, self-organization, the reduction principle, etc.). The reverse process of building a complex model by stage-by-stage complication of some initial set of simple models is also considered, as well as the deployment of a hierarchy of models of varying degrees of complexity.

There is a tendency to systematize knowledge in certain fields of science, or rather in interdisciplinary research. If earlier these were single books, review articles with a huge list of references, now there is a systematization of knowledge in the form of specialized encyclopedias. For example, Elsevier publishes materials on ScienceDirect and the list includes many "Reference Materials", which can be found at https://www.digento.de/titel/102260.html . On the topic of this work, a number of encyclopedias can be distinguished as additional literature - [207-212]. It is assumed that over time, such knowledge bases in certain areas will be made by artificial intelligence in order to more quickly form relevant knowledge.

**3. RESULTS AND DISCUSSION**.

The main weaknesses of the proposed multivariate approach are:

1) the forecasting model of the human thermoregulation system does not contain enough data from the open literature to predict its state due to the lack of experimental and theoretical data.

2) not revealed all the relationships, their nature in the form of functions (dependencies) in the proposed classification of factors of the functioning of the human thermoregulation system.

3) the classical and simple approach to the study of one of the systems of the human body presented in this work can be applied to other systems, living organisms.

4) there are other approaches to the study of multifactorial biological systems.

For example, molecular dynamics methods, genetic algorithms (biology) used to study the organism are not used for the embryo from the moment of its conception. In the near future, this is not possible due to the lack of the necessary specialists, legal framework and the presence of ethical standards for conducting such experiments on artificial organisms.

The proposed solution is not complete, since even in the classification itself there is not enough experimental and theoretical data in order to more correctly model both the thermoregulation system itself and the local biological processes of the body on the action of high or low temperatures at the molecular and cellular levels.

A feature of modern research in the field of sports medicine and disaster medicine is the lack of taking into account the individual characteristics of the organism. Sometimes the results are generalized, claiming new limits of the human body, although these are particular individual cases. From the point of view of thermophysical aspects and the functioning of the thermoregulation system of the human body, these features are studied in other sciences (medical physics, biophysics and biochemistry).

In the regulatory documents and documents on labor protection, there are also a number of restrictions associated with physical activity on a person, their duration and environmental parameters. These limits are derived from statistical processing of observed physiological experiments and health criteria. Taking into account the individual characteristics (not only thermophysical) of a person can more accurately determine labor productivity, whether it is possible to continue work or service in such conditions and for how long for each specific person.

In further studies of the individual human thermoregulation system, it is proposed to use systems of internal and external control of influencing factors. For internal control of the human body, a system with blood test sensors can be used as an example of an existing monitoring system. To monitor the environment and predict its changes, there are national and international meteorological services, according to which it is possible to determine, for example, the effective temperature UTCI or more specifically measured parameters of the environment affecting humans.

National systems of standards for assessing the comfortable state of a person for his work activity take into account statistical models and semi-empirical data that were obtained during the experiment, but generalized to populations. This is not the correct approach, even from the point of view of experiment planning, and the interpretation of the results for larger samples of people and consolidation in industry government standards is not acceptable at all. However, associations and NGOs came to the rescue, which began to study thermal comfort already according to different basic principles, relying more on taking into account climatic factors and a person's response to climate. The article proposes to go another way of forming standards for assessing the state of the body, taking into account the accumulated scientific experience in the field of thermoregulation and thermophysical aspects.

It is important to correctly generalize from private experiments to a general group of people in any country. At present, this is not observed, or the results of experiments are incorrectly generalized. An example would be technical standards from the industrial sector, designed to regulate the temperature on the surfaces of work units and mechanisms, with which a worker comes into contact with no means or with protective equipment. Another example would be standards for defining comfortable working conditions for workers. Although a huge number of experiments have been carried out and measurement methods have been created, they are all based on subjective sensations, statistical models, engineering empirical formulas that describe the system at a qualitative level, but the quantitative ratios that are given in the standards for ensuring the comfort of workers are already based on statistical samples and empirical observations. This approach can and should be developed from statistical and subjective sensations in the direction of general empirically described patterns, taking into account the individual characteristics of a person, which can be obtained experimentally and quickly.

## 4. CONCLUSIONS.

A lot of experimental data have been accumulated, methods for studying biological systems in medical physics, biophysics and biochemistry have appeared and developed, which make it possible to more accurately describe in a quantitative ratio and find new relationships in biological processes in the body, as well as the relationship between the systems of the human body.

There is no personalized method for predicting the functioning of the human thermoregulation system for medical purposes and in the field of labor protection. Existing approaches to protecting the health of workers are little applied and implemented, or are ignored in organizations.

To create a personalized predictive model of a human thermoregulation system, it is necessary to collect a lot of data in order to compose a model problem that simulates a real biological system at all levels of the problem description (cellular, tissue, organ).

For medical purposes, the created general methods for calculating the functioning of the thermoregulation system are not acceptable and cannot be a standardized measurement method, since the experiments were carried out with limitations in the considered simulated problems. This is the most important stumbling block to overcome in the future. It is proposed not to go deductively, but inductively in building a model and methodology for predicting the thermal state of a person in the environment.

The individual characteristics of the human thermoregulation system at the F [C3] level can already be basically diagnosed and calculated, but this is not enough. The main discoveries are concentrated in the field of research F [C4] in biophysics and biochemistry. At present, mathematical methods are sufficiently developed to study body systems at the subcellular and quantum level, thanks to methods from biophysics and quantum mechanics. Measuring technology is still far behind. The computing power of computers is already close to those that are necessary for calculating the systems of the human body at the cellular level, but so far they are busy with more urgent tasks and there are few programmers who can describe tasks for the systems of the body at the cellular level.

The proposed classification of tasks is at the same time a tool for systematizing scientific knowledge and searching for unexplored aspects related to the human thermoregulation system and local heat and mass transfer processes. And in the future, with a set of theoretical knowledge and a large array of experimental data, it can serve as a multivariate model for predicting a person's condition and be a standardized calculation method, including in the form of a software and hardware complex with an online calculator on the Internet.

The proposed approach to the classification of the study of the individual thermoregulation system makes it possible to form a new basis for the approach to the study of the national characteristics of the system of a person's comfort state, burn injuries and the response biological connection to external environmental influences. The described techniques and methods require further development both in technical and methodological terms, so that the mathematical model of the individual thermoregulation system can be applied to the tasks of medicine and labor protection.

## 4.1 Afterword. Possible stages in the development of modeling of a human thermoregulation system.

By developing a predictive model, it is possible to give ways of its development in the form of an increase in the number of dependent parameters and factors.

Stages of creating a forecasting system.

1) Statistical ideal model. The number of parameters i = 5 000.

2) Refined model based on scattered data. i = 30 000.

3) Dynamic screening individual model. $i = 10^8$. This is the tissue level. Based on experimental data on a specific person and a complete assessment of the state of the body.

4) Generalized tissue physiological model for medicine of the future. $i = 10^{20}$.

5) Biophysical individual model based on theory and new diagnostic techniques. This is at the cellular level. Suitable model for individual planning of pharmacological and other patient care. Very close to the real model, but described by an ensemble of cells, not by each cell.

6) Perhaps, in the study of embryos and artificial humans, the subcellular level of description of the biological system will be applied in the future. The description of biosystems will not become ensembles of cells, but a description of each cell and its development and reproduction in the process of evolution and division from the moment the embryo is conceived.

**Data availability**. The data supporting the findings of this study are publicly available on e-journal websites, indexed journals at http://doi.org, university website, Google Scholar (books) and are obtained from public domain resources.

This study did not receive any specific grant from funding agencies in the government, commercial or non-profit sectors.

**Acknowledgments.**

**Conflict of interest:** none.

## 5. REFERENCE. (Available for download https://cloud.mail.ru/public/HksN/WJfiL854q)